\documentclass[aps,pra,showpacs,twocolumn,superscriptaddress]{revtex4}
\usepackage[T1]{fontenc}
\usepackage[latin9]{inputenc}
\usepackage[english]{babel}
\usepackage{amstext}
\usepackage{graphicx,color}
\usepackage{float}
\usepackage{ulem}

\makeatletter
\@ifundefined{textcolor}{}
{%
 \definecolor{BLACK}{gray}{0}
 \definecolor{WHITE}{gray}{1}
 \definecolor{RED}{rgb}{1,0,0}
 \definecolor{GREEN}{rgb}{0,1,0}
 \definecolor{BLUE}{rgb}{0,0,1}
 \definecolor{CYAN}{cmyk}{1,0,0,0}
 \definecolor{MAGENTA}{cmyk}{0,1,0,0}
 \definecolor{YELLOW}{cmyk}{0,0,1,0}
}

\makeatother

\usepackage{babel}
\begin{document}

\newcommand{\red}[1]{\textcolor{red}{#1}}
\title{On the Monogamy of Entanglement of Formation}

\author{Thiago R. de Oliveira}
\affiliation{Instituto de F\'isica, Universidade Federal Fluminense, Av. Gal. Milton
Tavares de Souza s/n, Gragoat\'a, 24210-346, Niter\^oi, RJ, Brazil}

\author{Marcio F. Cornelio}
\affiliation{Instituto de  F\'\i sica, Universidade Federal de Mato Grosso, CEP 78060-900, Cuiab\'a, 
MT, Brazil}

\author{Felipe F. Fanchini}
\affiliation{Faculdade de Ci\^encias, UNESP - Universidade Estadual Paulista, CEP 17033-360, 
Bauru, SP, Brazil}

\begin{abstract}
It is well known that a particle cannot freely share entanglement with two or more particles.
This restriction is generally called monogamy. However the formal quantification of such
restriction is only known for some measures of entanglement and for two-level systems.
The first and broadly known monogamy
relation was established by Coffman, Kundu and Wootters for the square of the concurrence. Since
then, it is usually said that the entanglement of formation is not monogamous, as it does not
obey the same relation. We show here that despite that, the entanglement of formation can not be freely
shared and therefore should be said to be monogamous. Furthermore, the square of the entanglement of
formation does obey the same relation of the squared concurrence, a fact recently noted
for three particles and extend here for $N$ particles. Therefore the entanglement of formation is as
monogamous as the concurrence. We also numerically study how the entanglement
is distributed in pure states of three qubits and the relation between the sum of the bipartite
entanglement and the classical correlation. 
\end{abstract}

\maketitle

\section{Introduction}

In the last decades, we have seen many advances in the theory of entanglement 
\cite{HorodeckiReview} and, more generally, in the theory of quantum correlation \cite{rmp}. 
Nonetheless, the setting with more than two parts is still a challenge even from a 
conceptual point of view, not to mention the quantification. In this
scene monogamy relations are important, as they may indicate a structure for correlation in the 
multipartite setting. Monogamy has also been found to be the essential feature allowing for 
security in quantum key distribution \cite{Pawlowski}.
In the literature, the concept of monogamy for a entanglement measure $E$ becomes synonymous 
with $E$ satisfying the inequality
\begin{equation}
E_{1|23} \ge E_{12} + E_{13},
\label{monogamy}
\end{equation}
where $1$, $2$ and $3$ mean the respective parts of a tripartite system. 
However, most measures of entanglement do not satisfy inequality (\ref{monogamy}) with the 
exception of the squashed entanglement 
\cite{Koashi2004}.

The first monogamy relation established was due to Coffman, Kundu, and Wootters (CKW) \cite{Kundu}
for three qubits and latter generalized for $N$ qubits \cite{Osborne}.
It relates the squared concurrence between bipartitions as follows: 
\begin{equation}
C_{1|23...N}^{2}\geq C_{12}^{2}+C_{13}^{2}+...+C_{1N}^{2}.
\label{ckwnqubits}
\end{equation}
Such relation tells us that the one particle (particle 1) cannot freely
share entanglement with other qubits, thus the name monogamy. 
This monogamous relation has also been studied for other entanglement
measures \cite{Koashi2004,Cornelio2010,Bai} and also for the discord 
\cite{Fanchini11,Giorgi11,Prabhu12,Braga}. It is well known that the entanglement of formation,
when not squared, does not obey the inequality given by Eq. (\ref{monogamy}). Thus it
is usually said that the entanglement of formation is not monogamous.

In this work, we discuss the concept of monogamy beyond inequality given by Eq. (\ref{monogamy}), 
focusing on the entanglement of formation (EF). We argue 
that not satisfying the inequality given by Eq. (1) does not mean that a measure is 
not monogamous and can be freely shared. In fact, we numerically found an upper bound for
$E_{12} + E_{13}$ using the entanglement of formation,
which is considerably smaller than 2. Besides, very recently it was shown by 
Bai {\it et al} \cite{Bai} that, when squared, EF does satisfy a 
monogamy relation for three-qubit systems like the concurrence. We also generalize this result 
for $N$ qubits. Therefore there is no reason at all to say the the entanglement of formation
is not monogamous. Finally we analyze the distribution of the bipartite entanglement
in three-qubit pure states and the relation between it and the classical correlation.

The text is organized as follows: We begin by presenting a detailed study exploring
the distribution of quantum correlations, considering EF and concurrence. First, 
we study the squared EF for tripartite and multipartite systems. We point out that the 
EF is as monogamous exactly as the concurrence, since the squared EF does obey the CKW relation.
Following, we consider a system composed of three qubits and, numerically, obtain a
bound for $E_{12}+E_{13}$ and a similar result for the discord.
Moreover, we show how these monogamy relations behave for random states.

\section{Monogamy Relations}

The most basic and drastic monogamy relation happens for maximally
entangled pure states: two particles $S_1$ and $S_2$ maximally
entangled can not be entangled with a third particle. That happens
because to be maximally entangled they have to be in a pure state,
in an singlet, for example, $|\psi_{12}\rangle=(|01\rangle-|10\rangle)/\sqrt{2}$.
But a pure state has null entropy and therefore null correlation with any other system:
$S_{12}$  cannot be entangled, or even classically correlated, with a third
particle and thus neither can $S_1$ or $S_2$. However, this argument only applies
when the first two parts are in a pure state.
When they are in a mixed state the restriction is not so drastic: one particle may be entangled with two others at the same time. 

We divide our work in two parts: first we study the monogamy considering squared measures 
of entanglement and, in sequence, we explore the monogamy for each measure {\it per se}, i.e.,
the measure up to 1.

\subsection{Monogamy for the squared version}

\begin{figure}
\includegraphics[width=0.45\textwidth]{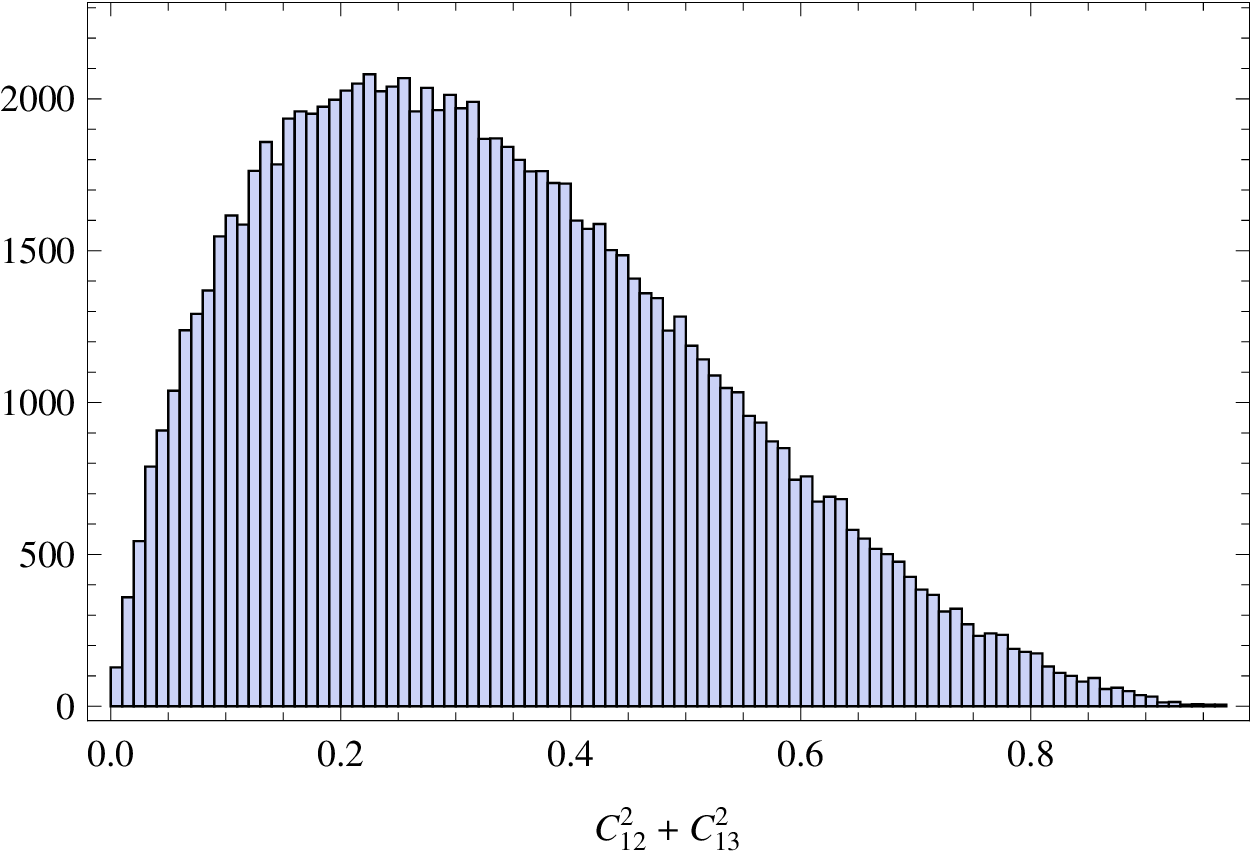}
\caption{\label{fig:Conc2}Sum of the square of the concurrence , $C_{12}^{2}+C_{13}^{2}$
for random pure states of three qubits using the Haar measure.}
\end{figure}

For three qubits, the amount of total entanglement
that can be shared is restricted by the CKW inequality:
\begin{equation}
C_{12}^{2}+C_{13}^{2}\leq C_{1|23}^{2}\leq1.
\label{ckw}
\end{equation}
Note that here $C_{1|23}$ is the concurrence between qubit
$1$ and the joint qubits $23$, which can be analytically
obtained for pure states and it is at most $1$. In Fig. \ref{fig:Conc2} we
show a histogram of the value of $C_{12}^{2}+C_{13}^{2}$
for random pure states of three qubits sampled uniformly (Haar measure 
\footnote{All the figures were made using $10^5$ random states to make the files smaller. But we checked that the same behavior holds using $10^6$ random states} ).
It can be seen that few states are close to the upper bound $1$. As mentioned
before the monogamy relation is also true for $N$ particles as shown in
\cite{Osborne}. 

Concurrence was actually introduced as an intermediate measure to obtain
the entanglement of formation, but, as it is a monotonic function of
EF, it is usually used as the entanglement measure \cite{Bennett96,Wooters98}.
However, contrary to EF, it has no clear operational meaning. Therefore, it is
natural to ask whether the EF also obeys the relation above. Already in
the original paper, CKW have shown that EF does not obey the monogamy
relation, since it is a concave function of $C^{2}$. It is important to emphasize, however, that the 
authors even mention that this is not a paradox, since they have only shown that
EF does not obey this particular kind of monogamy relation, given by Eq. (\ref{ckw}).
Since then, it is usually said that EF is not monogamous. It is curious
that the authors analyzed EF and not its square, since they were working
with $C^{2}$ and not $C$. 

In fact it is easy to show that $E_F^{2}$
does obey the monogamy relation for qubits, and this fact was first noticed by Bai {\it et al} \cite{Bai}.
Their proof is for 3 qubits, but their argument is equally valid for $N$ qubits.
Actually, in the first paper of CKW \cite{Kundu}, it was already noticed that any monotonic convex function of $C^2$ would also be a monogamous measure of entanglement.
Indeed, that is exactly the difference between the EF and the squared EF: EF is a concave function of $C^2$ while the squared EF is a convex function 
of $C^2$. Given that the general proof cannot be found either in Ref. \cite{Kundu} or in Ref. \cite{Bai}, we reproduce their arguments here for the general case. We start with the fact that the $E_F^2$ is a monotonic function of $C^2$ and, using the CKW monogamy for $N$ qubits, we have
\[
E_F^2\left(C^2_{1|23...N}\right) \ge E_F^2\left(\sum_{i=2}^N C^2_{1|i}\right).
\]
Now we consider the expression
\begin{equation}
\tau_{E_F} = E_F^2\left(\sum_{i=2}^N C^2_{1|i}\right) - \sum_{i=2}^N E_F^2(C_{1|i}^2).
\label{tangleEOF2} 
\end{equation}
We have to prove that it is greater than zero. For that, we use the fact that $E_F^2$ is a convex function of $C^2$. This implies that the inclination $m$ of the straight lines from the origen to the points $\{E_F^2(C^2),C^2\}$ increases monotonically with $C^2$. That is, we have, for
\[
m = \frac{E_F^2\left(\sum_{i=2}^N C^2_{1|i}\right)}{\sum_{i=2}^N C^2_{1|i}}
\hspace{.5cm} \textrm{and} \hspace{.5cm}
m_i = \frac{E_F^2(C^2_{1|i})}{C^2_{1|i}}
\]  
that $ m \ge m_i$ for all $i$. Replacing the expressions for the $m$'s in Eq. (\ref{tangleEOF2}) and using that $m \ge m_i$ we have that $\tau_{E_F}$ is always positive.

\begin{figure}
\includegraphics[width=0.45\textwidth]{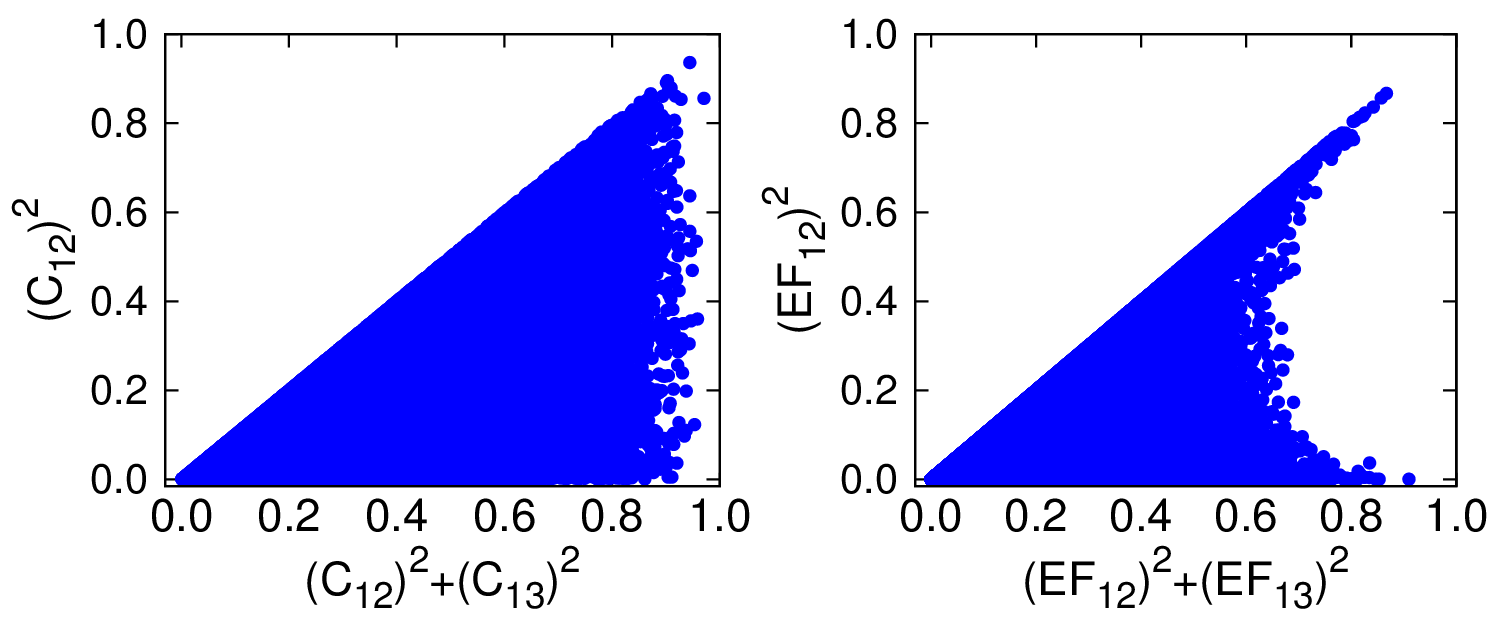}
\caption{\label{fig:Distribution}Distribution of the square of the entanglement between
the three particles for random pure states of three qubits using the Haar
measure. One can see that the saturation of the bound for the
square of the concurrence can come just from one pair of particles
or from both. In the case of the square of entanglement of formation
it comes exclusively from one of the pairs.}
\end{figure}

Following, we also study how the entanglement is distributed in the states. More
specifically, we are interested in the following question: do the
states which are close to the bound have most of their entanglement
from one pair and just a little between the other pair or does it
comes from both pairs? In Fig. \ref{fig:Distribution} we plot the sum of
the square of the entanglement versus the entanglement of one the
pairs. We can see that, for the concurrence, it seems that the saturation
of the bound can come just from the entanglement of one pair, or
from both pairs. On the other hand, the saturation of the square of
EF comes exclusively from one of the pairs: there are no state close
to the bound with $E_{12}^{2}\approx E_{13}^{2}$. This shows that
although monotonically related the concurrence and EF may have different
qualitative properties. 

\subsection{Monogamy for the linear version}

\begin{figure}
\includegraphics[width=0.45\textwidth]{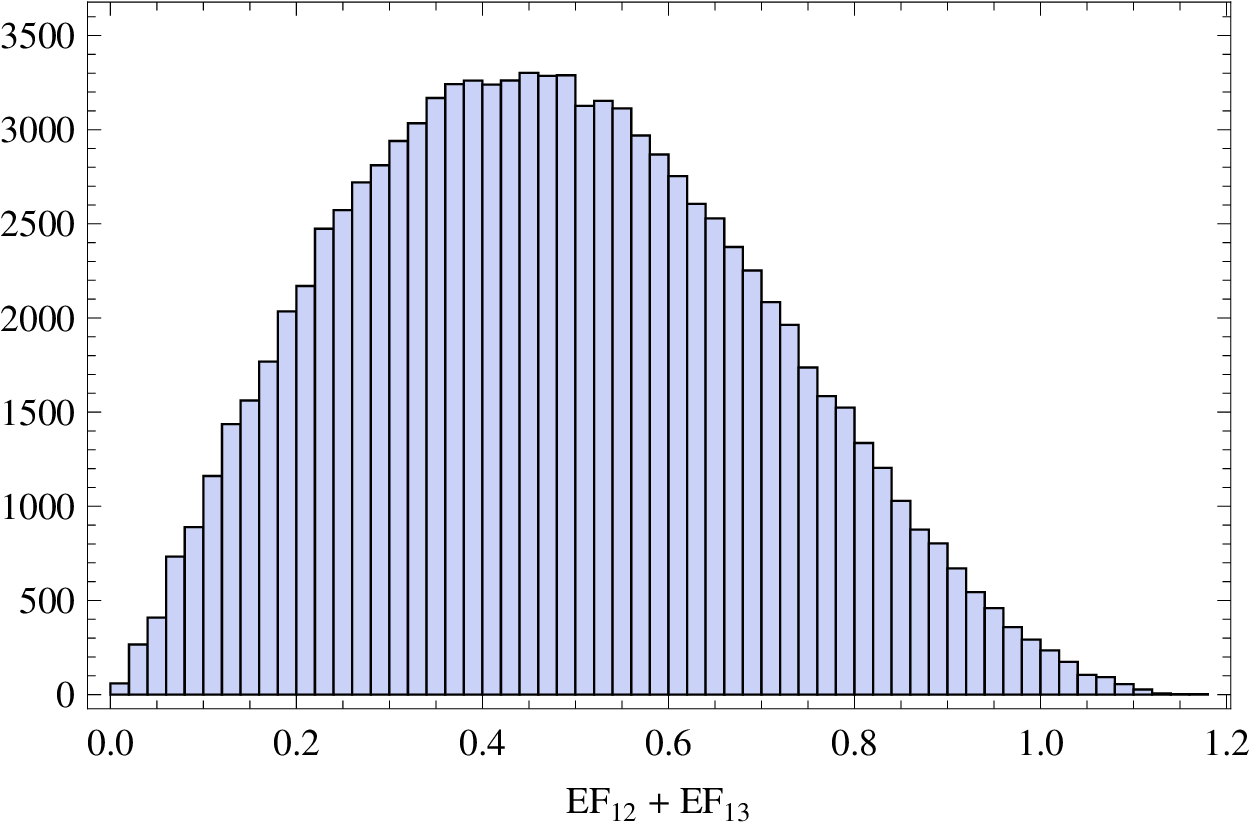}
\caption{\label{fig:EF}Sum of the entanglement of formation , $E_{12}+E_{13}$
for random pure states of three qubits using the Haar measure.}
\end{figure}

\begin{figure}
\includegraphics[width=0.45\textwidth]{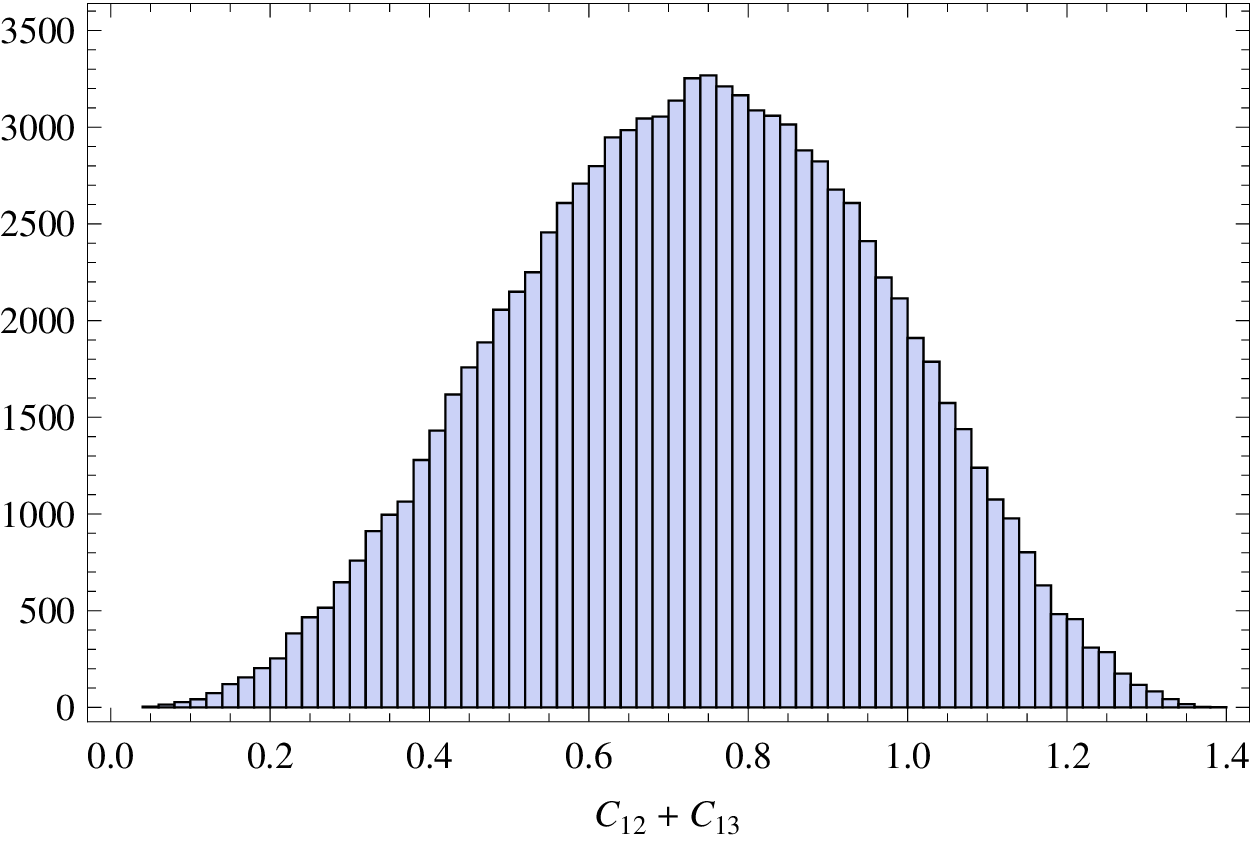}
\caption{\label{fig:Conc}Sum of the of the concurrence, $C_{12}+C_{13}$
for random pure states of three qubits using the Haar measure.}
\end{figure}

\begin{figure}
\includegraphics[width=0.45\textwidth]{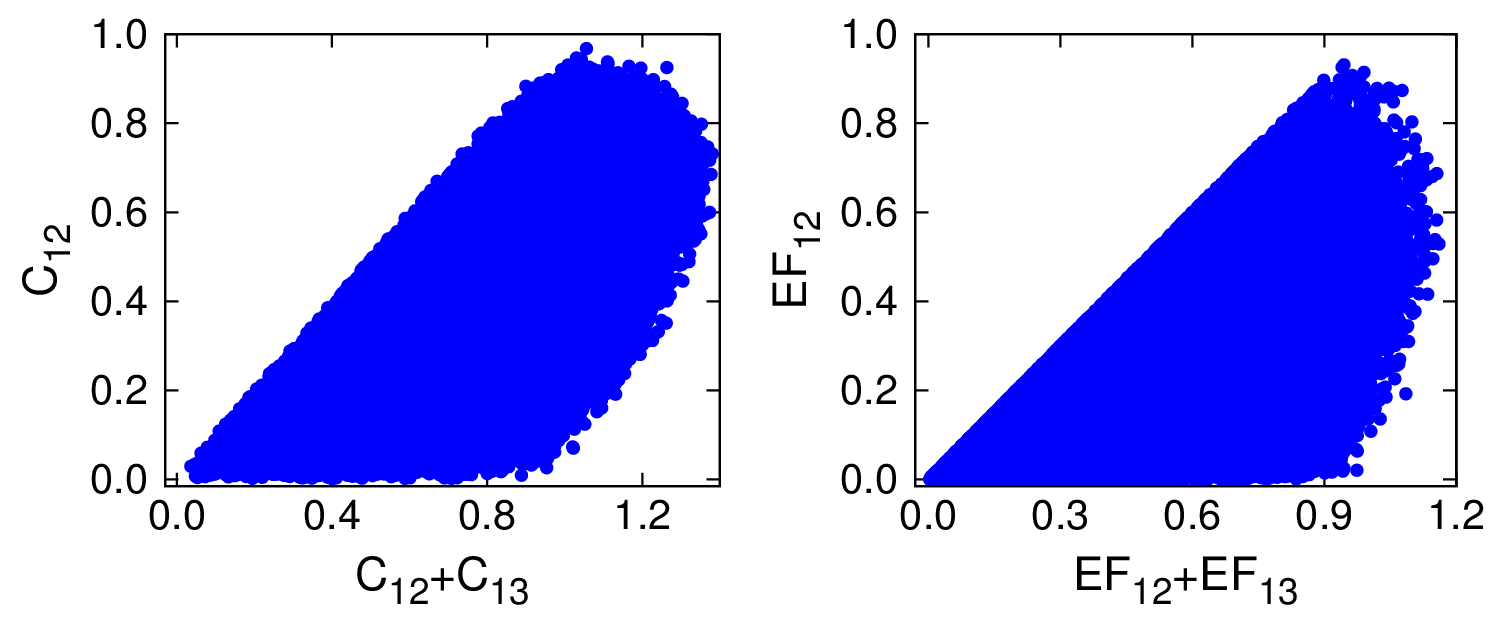}
\caption{\label{fig:DistributionConcEF} Distribution of entanglement between the
three particles for random pure states of three qubits using the Haar
measure. One can see that, contrary to the case for the square of the entanglement,
there is no visible difference between the behavior of the concurrence and the entanglement of 
formation.}
\end{figure}

Noting that $E_F$ does not obey the monogamy relation given by Eq. (\ref{ckw}),
should we say it is not monogamous? Can it be freely shared between
three particles? This is the question we address here. We want to
find an upper bound for
\[
E_{12}+E_{13}.
\]
We know the sum cannot achieve 2, but how close can it be? We studied
this case numerically for a system of three qubits considering a sampling of $10^6$ random states uniformly 
distributed and we found that
\[
E_{12}+E_{13}\leq 1.18819
\]
In Fig. \ref{fig:EF} we plot the value of the sum of EF for random
states sampled uniformly. We can note that there is an upper bound and that just a few
states are close to it. Thus it is at least misleading to say that the EF is not
monogamous. Indeed, as we can see in Fig. \ref{fig:EF}, there are strong constrains on how it can be shared. Furthermore, even concurrence also does not obey the usual monogamy relation
as can be seen in Fig. \ref{fig:Conc}. We also study the distribution of the entanglement, as we did for the square of the measures, in Fig \ref{fig:DistributionConcEF}. We can see that in this case the behaviors of the concurrence and the EF are similar. This indicates that the square of the measure can have different qualitative behavior from the measure itself.



We should also point out that for pure states of three qubits there
is a conservation law between discord and EF, such that $E_{12}+E_{13}=D_{12}+D_{13}$
\cite{Fanchini11}. So the monogamies of the two quantum correlations
are tightly connected, and a violation of one would imply of the other,
something already noted before by some of us \cite{Fanchini13}. 
These relations are obtained from the Koashi-Winter (KW) relation $E_{12}+J_{13}^\leftarrow=S_{1}
$ \cite{Koashi2004},
where $J_{13}^\leftarrow$ is the one way classical correlation between
$1$ and $3$ \cite{Henderson01,Olivier01}. The classical correlation is defined as the condition 
entropy after measurements:
$J_{13}^\leftarrow=\textrm{max}_{\Pi_{x}^{3}}[S(\rho_{1})-\sum_{x}p_{x}S(\rho_{1}^{x})]$
with $\rho_{1}^{x}$ being the reduced state of $1$, after a measurement
made on $3$ with $x$ as the result. Additionally, a maximization over
all measurements $\Pi_{x}^{3}$ on $3$ is performed. 

\begin{figure}
\includegraphics[width=0.45\textwidth]{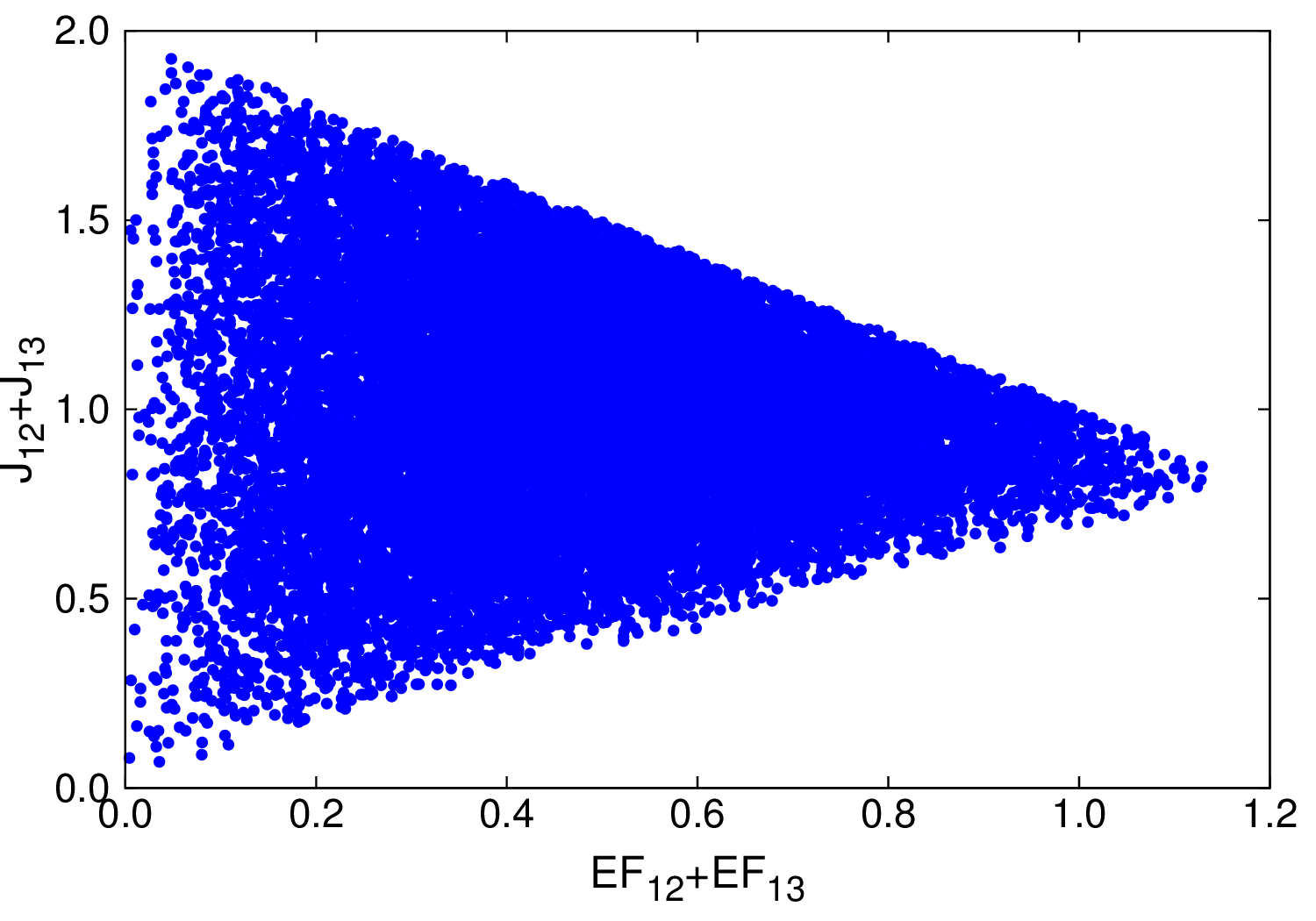}
\caption{\label{fig:SUMJSUMEF}
Relation betweem $J_{12}^\leftarrow+J_{13}^\leftarrow$
and $E_{12}+E_{13}$ for random pure state of three qubits using the
Haar measure.}\label{Classical}
\end{figure}

Is it also worth noting that using such relation, we can obtain some insight about the state that maximizes $E_{12}+E_{13}$.  The idea is to use the KW relation to write the bipartite EF sum as a function of the entropy $S_1$. Our point is to answer an important question: what is the state 
and the value of entanglement between subsystems $1$ and $23$ that saturates the monogamy inequality? Is it a maximally entangled state? To answer this question we begin with an expression where the sum of the bipartite EF is written as a function of the sum of the bipartite classical correlation and the entropy $S_1$. We use the KW relation twice to obtain that $J_{12}^\leftarrow +J_{13}^\leftarrow=2S_{1}-(E_{12}+E_{13})$. The sum of the classical correlation as a function of the sum of the EF is plotted in Fig. \ref{fig:SUMJSUMEF}.
It can be seen that, as the states increase the sum of the EF, the range of possible values of the classical correlation decrease and goes 
to around 0.8. At this situation we note that the sum of EF tends to $1.2$ since $J_{12}^\leftarrow +J_{13}^\leftarrow+(E_{12}+E_{13})=2S_{1}\approx2$. It means that the maximum value of $E_{12}+E_{13}$ occurs when $S_1=1$, i.e. when subsystem $1$ is maximally entangled with subsystem $23$. With this result we then try to
 determine the state that maximizes $E_{12}+E_{13}$. Since we look for a state with $S_1=1$ and the sum of bipartite entanglement maximum, we discard the GHZ states (since for these states $E_{12}=E_{13}=0$) which leads us to the $W$ state given by
\begin{equation}
|\Psi\rangle=\sqrt{\frac{1}{2}}|100\rangle + \frac{1}{2}\left(|010\rangle + |001\rangle\right).
\end{equation}
This state has $S_1=1$ and the sum of the bipartite EF is give by $1.20175$. This result is greater than all of our numerical results which strongly suggests that the saturation of the monogamy inequality is reached when subsystem $1$ is maximally entangled with subsytem $23$.

\section{conclusion}
 
We have studied how the entanglement between three particles may be shared.
We numerically found un upper bound on the sum of the entanglement of formation,
$E_{12}+E_{13}$, showing that it can not be freely shared even though it does not
obey the Coffman, Kundu, and Wootters (CKW) relation $E_{1|23} \ge E_{12} + E_{13}$.
Furthermore, the square of the
entanglement of formation does obey the relation, as shown here for $N$ particles
extending the proof for three particles \cite{Bai}. Thus the entanglement of formation is
as monogamous as the concurrence: the square of both obeys the relation, but
not the measure itself which is limited by another upper bound that we numerically found.
Interestingly, the states with maximum sum of the square of the entanglement of formation
have their entanglement coming mostly from one of the pairs, while for the concurrence
it can come from both pairs. Finally we analyzed the relation between the sum of
the entanglement of formation and the classical correlation.

As a perspective one could define a genuine tripartite entanglement measure analogous
to the tangle but using the square of the entanglement of formation instead of the square of 
the concurrence $\tau_F= E_F^2(1|23)-E_F^2(1|2)-E_F^2(1|3)$. We already observed that $\tau_F$
is 1 for the GHZ state, but not null for the W state. Note that when using the concurrence $\tau$ is null for the W state and this is usually used to argue that the W state has no genuine tripartite 
entanglement. So one could argue that the W does have genuine tripartite entanglement or that $\tau$
are not really good measures of genuine tripartite entanglement. But before that one should first
study $\tau_F$ with more care to see if it is really a {\it bona fide} entanglement measure: for example, is it
monotonic under local operation and classical communications (LOCC)?

 
 

\emph{Note added} - After submission of this manuscript we became aware
of related work, arXiv:1401.3205, in which the monogamy inequality
is proved also for mixed states.

\emph{Acknowledgements} - We would like to thanks the organizers of "IV Quantum Information School and Workshop of Paraty" where the discussions which lead to this work were started in the very informal and fruitfulbeach atmosphere of the workshop. We are also in debt to Y. K. Bai for clarifying discussions about monogamy of the squared entanglement of formation. The authors are supported by the National Institute for Science and Technology of Quantum Information (INCT-IQ) under process number 2008/57856-6 and FFF is supported by S\~{a}o Paulo Research Foundation (FAPESP) under grant number 2012/50464-0, and by the National Counsel of Technological and Scientific Development (CNPq) under grant number 474592/2013-8.

\end{document}